\documentclass[aps,pra,english,twocolumn,notitlepage,showpacs,nofootinbib]{revtex4-1}
\usepackage{mathptmx}

\usepackage[T1]{fontenc}
\usepackage[latin9]{inputenc}
\usepackage{color}
\usepackage{babel}
\usepackage{textcomp}
\usepackage{bm}
\usepackage{graphicx}
\usepackage{amsmath}
\usepackage{amsthm}
\usepackage{amssymb}
\usepackage{epstopdf}
\usepackage[unicode=true,pdfusetitle,
bookmarks=true,bookmarksnumbered=false,bookmarksopen=false,
breaklinks=true,pdfborder={0 0 0},pdfborderstyle={},backref=false,colorlinks=true]
{hyperref}
\hypersetup{
	urlcolor=blue, citecolor=blue, linkcolor=blue}

\makeatletter
\usepackage{amsmath, amsfonts, amssymb, amsthm, mathrsfs, amssymb, braket, bbm, bm,graphicx,times,txfonts,color,subfigure}
\usepackage{hyperref}

\theoremstyle{plain}
\newtheorem{Theorem}{Theorem}

\newtheorem{Example}{Example}
\newtheorem{Lemma}{Lemma}
\newtheorem{Definition}{Definition}
\newtheorem{Corollary}{Corollary}

\makeatother

\begin{document}

\title{Coherence and imaginarity of quantum states}

\date{\today}

\title{Coherence and imaginarity of quantum states}
\author{Jianwei Xu}
\email{xxujianwei@nwafu.edu.cn}
\affiliation{College of Science, Northwest A$\&$F University, Yangling, Shaanxi 712100,
China}

\begin{abstract}
Baumgratz, Cramer and Plenio established a rigorous framework (BCP
framework) for quantifying the coherence of quantum states [\href{http://dx.doi.org/10.1103/PhysRevLett.113.140401}{Phys. Rev. Lett.
113, 140401 (2014)}]. In BCP framework, a quantum state is called incoherent
if it is diagonal in the fixed orthonormal basis, and a coherence measure
should satisfy some conditions. For a fixed orthonormal basis, if a quantum
state $\rho $ has nonzero imaginary part, then
$\rho $ must be coherent. How to quantitatively characterize this fact? In
this work, we show that any coherence measure $C$ in BCP framework has the
property $C(\rho )-C($Re$\rho )\geq 0$ if $C$ is invariant under state
complex conjugation, i.e., $C(\rho )=C(\rho ^{\ast })$, here $\rho ^{\ast }$
is the conjugate of $\rho ,$ Re$\rho $ is the real part of $\rho .$ If $C$
does not satisfy $C(\rho )=C(\rho ^{\ast }),$ we can define a new coherence
measure $C^{\prime }(\rho )=\frac{1}{2}[C(\rho )+C(\rho ^{\ast })]$ such
that $C^{\prime }(\rho )=C^{\prime }(\rho ^{\ast }).$ We also establish some similar results for bosonic Gaussian
states.
\end{abstract}
\maketitle

\section{Introduction}
Coherence is a key feature of quantum states. In 2014, Baumgratz, Cramer and
Plenio established a rigorous framework (BCP framework) for quantifying the
coherence of quantum states \cite{PRL-Plenio-2014}. Although there have
been fruitful results about coherence in both theories and experiments under
BCP framework \cite{RMP-Plenio-2017,Fan-2018-PhysicsReports,AQT-2021-Wu}, coherence is still under
active research.

In BCP framework, coherence is basis dependent, we suppose the fixed
orthonormal basis is $\{|j\rangle \}_{j=1}^{d},$ here $d$ is the dimension
of complex Hilbert space $\mathbb{C}^{d}$ associating to the quantum system under study. A quantum state can be
represented by a density operator $\rho .$ Expanding $\rho $ in the basis $%
\{|j\rangle \}_{j=1}^{d}$ leads to
\begin{equation}
\rho =\sum_{j,k=1}^{d}\rho _{jk}|j\rangle \langle k|     \label{eq1-1}
\end{equation}%
where $\rho _{jk}=\langle j|\rho |k\rangle .$ The state $\rho $ is called
incoherent if $\rho _{jk}=0$ for all $j\neq k,$ i.e., $\rho $ is diagonal in
$\{|j\rangle \}_{j=1}^{d}.$ $\rho $ is called coherent if $\rho $ has at
least one nonzero off-diagonal element. A quantum operation  \cite{Nielsen-2010-quantum} $\phi $ is often
expressed in the form of Kraus operators $\phi =\{K_{\mu }\}_{\mu },$ where
all $K_{\mu }$ are operators on $\mathbb{C}^{d}$ and satisfy $\sum_{\mu }K_{\mu }\leq I_{d},$ here $I_{d}$ is the
identity operator of dimension $d,$ $\sum_{\mu }K_{\mu }\leq I_{d}$ means
that $I_{d}-\sum_{\mu }K_{\mu }\geq 0,$ i.e., $I_{d}-\sum_{\mu }K_{\mu }$ is
positive semidefinite. A quantum operation $\phi $ is called a quantum
channel if $\sum_{\mu }K_{\mu }=I_{d}.$ In BCP framework, a quantum
operation $\phi =\{K_{\mu }\}_{\mu }$ is called incoherent if $K_{\mu }\rho
K_{\mu }^{\dagger }$ is diagonal for all incoherent state $\rho $ and all $%
\mu .$ We use $K_{\mu }^{\ast },$ $K_{\mu }^{T}$ and $K_{\mu }^{\dagger }$
to denote the (complex) conjugate, transpose and conjugate transpose of $%
K_{\mu },$ respectively. A real valued functional $C$ defined on density
operators is called a coherence measure if $C$ satisfies the following
(C1)-(C4).

\textbf{(C1)} Faithfulness. $C(\rho )\geq 0$ for any state $\rho ,$ and $C(\rho )=0$
if and only if $\rho $ is diagonal.

\textbf{(C2)} Monotonicity. $C(\sum_{\mu }K_{\mu }\rho K_{\mu }^{\dagger })\leq
C(\rho )$ for any incoherent channel $\phi =\{K_{\mu }\}_{\mu }.$

\textbf{(C3)} Probabilistic monotonicity. $\sum_{\mu }C\left[ \frac{K_{\mu }\rho
K_{\mu }^{\dagger }}{\text{tr}(K_{\mu }\rho K_{\mu }^{\dagger })}\right]
\leq C(\rho )$ for any incoherent channel $\phi =\{K_{\mu }\}_{\mu }.$

\textbf{(C4)} Convexity. $C(\sum_{\mu }p_{\mu }\rho _{\mu })\leq \sum_{\mu }p_{\mu
}C(\rho _{\mu })$ for any probability distribution $\{p_{\mu }\}_{\mu }$ and
any states $\{\rho _{\mu }\}_{\mu }.$

Note that (C3) and (C4) imply (C2). Condition (C5) was proposed in Ref. \cite{PRA-2016-Tong} as follows.

\textbf{(C5)} Additivity for direct sum states.
\begin{equation}
C[p\rho _{1}\oplus (1-p)\rho _{2}]=pC(\rho _{1})\oplus (1-p)C(\rho _{2}),    \label{eq1-2}
\end{equation}
with $p\in \lbrack 0,1],$ $\rho _{1}$, $\rho _{2}$ any states. It is
shown that \cite{PRA-2016-Tong} (C2) and (C5) are equivalent to (C3) and (C4).

Roughly speaking, coherence theory characterizes how much the off-diagonal
part of a quantum state. Another feature of quantum states, imaginarity,
recently receives much attention \cite{JPA-Gour-2018,PRL-Guo-2021,PRA-Guo-2021,PRAP-2021-Guo,QIP-2021-Li,Nature-2021-Acin,PRR-2021-Zhu,PRL-2022-Li,NJP-2022-Streltsov,PRL-2023-Streltsov,PRA-2023-Xu,arXiv-2023-Guo}. Under the fixed orthonormal basis
$\{|j\rangle \}_{j=1}^{d}$ which is the same as in coherence theory, we
write the state $\rho $ in the form
\begin{equation}
\rho =\text{Re}\rho +i\text{Im}\rho ,      \label{eq1-3}
\end{equation}%
where $i=\sqrt{-1}$, $\text{Re}\rho =\sum_{j,k=1}^{d}($Re$\rho _{jk})|j\rangle \langle k|$
is the real part of $\rho ,$ Im$\rho =\sum_{j,k=1}^{d}($Im$\rho
_{jk})|j\rangle \langle k|$ is the imaginary part of $\rho .$ We say that
state $\rho $ is real if Im$\rho =0,$ otherwise we say that $\rho $ has
imaginarity. Roughly speaking, imaginarity theory characterizes how much the
imaginary part of a quantum state. Notice that, imaginarity also depends on
the choice of the fixed orthonormal basis $\{|j\rangle \}_{j=1}^{d}.$ Both coherence theory and imaginarity theory can be viewed as special quantum resource theories \cite{IJMPB-Horodechi-2013,RMP-Gour-2019}.

Observe that, if $\text{Im}\rho \neq 0$ then $\rho $ must be coherent. That
is to say, imaginarity must imply coherence. Since Re$\rho $ is still a
quantum state, then for any coherence measure $C,$ the coherence of Re$\rho ,
$ $C($Re$\rho ),$ is well defined. A natural question then arises that for a
coherence measure $C$, whether $C(\rho )-C($Re$\rho )\geq 0$ holds for all
states? $C(\rho )-C($Re$\rho )$ quantitatively characterizes the fact that
imaginarity must imply coherence. In this work, we investigate whether $%
C(\rho )-C($Re$\rho )\geq 0$.

The remainder of this paper is organized as follows. In section \ref{Section-II}, we study
$C(\rho )-C($Re$\rho )$ for any quantum states and mainly focus on the case
of finite dimensions. In section \ref{Section-III}, we study the case of bosonic Gaussian
states. Section \ref{Section-IV} is a brief summary. We put some necessary details about symplectic spectrums in \hyperlink{Appendix}{Appendix}. 

\section{$C(\rho )-C(\text{Re}\rho)$ for any quantum state}  \label{Section-II}

\subsection{Invariance of coherence under state complex conjugation}

For a coherence measure $C,$ we consider the condition (C6) below.

\textbf{(C6)} Invariance under state complex conjugation.
\begin{equation}
C(\rho )=C(\rho ^{\ast })      \label{eq2A-1}
\end{equation}
for any state $\rho .$

Under (C6), we have Theorem \ref{Theorem-1} below.
\begin{Theorem} \label{Theorem-1}
Suppose the coherence measure $C$ is invariant under state
complex conjugation, then $C(\rho )-C($Re$\rho )\geq 0.$   
\end{Theorem}
\emph{Proof.} By (C4) and (C6) we have
\begin{equation*}
C(\text{Re}\rho )=C\left(\frac{\rho +\rho ^{\ast }}{2}\right)\leq \frac{1}{2}[C(\rho
)+C(\rho ^{\ast })]=C(\rho ).
\end{equation*}
$\hfill\square$

We do not know whether any coherence measure $C$ must satisfy $(C6).$ If
 a coherence measure $C$ violates $(C6),$ then Theorem \ref{Theorem-2} below
shows that we still have a method to define a new coherence measure $%
C^{\prime }$ such that $C^{\prime }$ satisfies $(C6).$

\begin{Theorem}    \label{Theorem-2}
For the coherence measure $C,$
\begin{equation}
C^{\prime }(\rho )=\frac{1}{2}[C(\rho )+C(\rho ^{\ast })]     \label{eq2A-2}
\end{equation}
is still a coherence measure and $C^{\prime }$ satisfies $(C6).$
\end{Theorem}

Before proving Theorem \ref{Theorem-2}, we give the definition of the conjugate of a
quantum operation.

\begin{Definition}   \label{Definition-1}
For a quantum operation $\phi =\{K_{\mu }\}_{\mu },$ we
define the conjugate of $\phi =\{K_{\mu }\}_{\mu }$ as $\phi ^{\ast
}=\{K_{\mu }^{\ast }\}_{\mu }.$
\end{Definition}

Since $\phi =\{K_{\mu }\}_{\mu }$ is a quantum operation, then $\sum_{\mu
}K_{\mu }^{\dagger }K_{\mu }\leq I_{d}.$ This ensures $\sum_{\mu }K_{\mu
}^{T}K_{\mu }^{\ast }\leq I_{d}$, that is to say, $\phi ^{\ast }=\{K_{\mu
}^{\ast }\}_{\mu }$ is still a quantum operation. Further, if $%
\phi =\{K_{\mu }\}_{\mu }$ is a quantum channel, that is, $\sum_{\mu }K_{\mu
}^{\dagger }K_{\mu }=I_{d},$ then $\sum_{\mu }K_{\mu }^{T}K_{\mu }^{\ast
}=I_{d},$ i.e., $\phi ^{\ast }=\{K_{\mu }^{\ast }\}_{\mu }$ is still a
quantum channel. We can directly verify that $\phi =\{K_{\mu }\}_{\mu }$ and
$\phi ^{\ast }=\{K_{\mu }^{\ast }\}_{\mu }$ have the property
\begin{equation}
\phi ^{\ast }(\rho ^{\ast })=[\phi (\rho )]^{\ast }       \label{eq2A-3}
\end{equation}%
for any state $\rho .$ Evidently, $\phi =\phi ^{\ast }$ if and only if $%
K_{\mu }=K_{\mu }^{\ast }$ for all $\mu ,$ namely, all $K_{\mu }$ are real
matrices. It is shown that $\phi =\{K_{\mu }\}_{\mu }$ is incoherent if and
only if for any $\mu ,$ each column of $K_{\mu }$ has at most one nonzero
element \cite{PRA-Du-2015}. This implies that if $\phi =\{K_{\mu }\}_{\mu }$ is
incoherent, then $\phi ^{\ast }=\{K_{\mu }^{\ast }\}_{\mu }$ is still
incoherent.

We now give a proof for Theorem \ref{Theorem-2}.

\emph{Proof of Theorem \ref{Theorem-2}.}
We prove that $C^{\prime }(\rho )$ satisfies (C1), (C2) and (C5). $C^{\prime
}(\rho )$ satisfying (C1) and (C5) can be easily checked, we only prove that
$C^{\prime }(\rho )$ satisfies (C2). For any incoherent operation $\phi
=\{K_{\mu }\}_{\mu },$ one has
\begin{eqnarray*}
C^{\prime }[\phi (\rho )] &=&\frac{1}{2}[C(\phi (\rho ))+C([\phi (\rho
)]^{\ast })] \\
&=&\frac{1}{2}[C(\phi (\rho ))+C(\phi ^{\ast }(\rho ^{\ast }))] \\
&\leq &\frac{1}{2}[C(\rho )+C(\rho ^{\ast })]=C^{\prime }(\rho ),
\end{eqnarray*}
where we have used $\phi ^{\ast }(\rho ^{\ast })=[\phi (\rho )]^{\ast }$,
the fact that $\phi ^{\ast }$ is incoherent, and (C2) that $C(\phi (\rho
))\leq C(\rho )$ and $C(\phi ^{\ast }(\rho ^{\ast }))\leq C(\rho ^{\ast }).$
$\hfill\square$

\subsection{Coherence measures satisfying (C6)}
With Theorem \ref{Theorem-1} and Theorem \ref{Theorem-2}, it is desirable to investigate whether the
existing coherence measures satisfy (C6). In this section, we address this
topic. Some coherence measures have been found under BCP framework as
follows. The $l_{1}$ norm of coherence is defined as \cite{PRL-Plenio-2014}
\begin{equation}
C_{l_{1}}(\rho )=\sum_{j\neq k}|\rho _{jk}|.      \label{eq2B-1}
\end{equation}
The relative entropy of coherence is defined as \cite{PRL-Plenio-2014}
\begin{equation}
C_{\text{r}}(\rho )=S(\rho _{\text{diag}})-S(\rho ),    \label{eq2B-2}
\end{equation}
where $\rho _{\text{diag}}=\sum_{j=1}^{d}\rho _{jj}|j\rangle \langle j|$ is
the diagonal part of $\rho ,$ $S(\rho )=-$tr$(\rho \log _{2}\rho )$ is the
von Neumann entropy of $\rho .$ The coherence measure based on Tsallis
relative entropy is defined as \cite{PRA-2017-Yu,SR-2018-Yu}
\begin{equation}
C_{\text{T},\alpha }(\rho )=\frac{1}{\alpha -1}\left[\sum_{j=1}^{d}\langle j|\rho
^{\alpha }|j\rangle^{1/\alpha } -1\right],\alpha \in \lbrack 0,1)\cup (1,2].     \label{eq2B-3}
\end{equation}
The robustness of coherence is defined as \cite{PRL-2016-Adesso,PRA-2016-Adesso}
\begin{equation}
C_{\text{R}}(\rho )=\min_{\tau }\left\{ s\geq 0 \Big|\frac{\rho +s\tau }{1+s}%
\text{ incoherent}\right\} ,      \label{eq2B-4}
\end{equation}
where $\min $ runs over all quantum states $\tau .$
The geometric coherence is defined as \cite{PRL-2015-Adesso}
\begin{equation}
C_{\text{g}}(\rho )=1-\max_{\sigma }[F(\rho ,\sigma )]^{2},     \label{eq2B-5}
\end{equation}
where $F(\rho ,\sigma )=$tr$\sqrt{\sqrt{\rho }\sigma \sqrt{\rho }}$ is the
quantum fidelity of state $\rho $ and $\sigma $, $\max $ runs over all
incoherent states $\sigma .$
The modified trace norm of coherence is defined as  \cite{PRA-2016-Tong}
\begin{equation}
C_{\text{tr}}(\rho )=\min_{\lambda \geq 0,\sigma }||\rho -\lambda \sigma ||_{%
\text{tr}},     \label{eq2B-6}
\end{equation}
where $||\rho -\lambda \sigma ||_{\text{tr}}$ is the trace norm of $\rho
-\lambda \sigma ,$ $\min $ runs over all $\lambda \geq 0$ and all incoherent
states $\sigma .$
The coherence weight is defined as \cite{PRA-2016-Bu}
\begin{equation}
C_{\text{w}}(\rho )=\min_{\sigma }\{s\geq 0:\rho \geq (1-s)\sigma \},    \label{eq2B-7}
\end{equation}
where $\min $ runs over all incoherent states $\sigma .$

Convex roof construction \cite{PRA-2015-Yuan,PRA-Du-2015,Qi-2015-QIC,JPA-2017-Yan,JPA-2022-Xu} provides a class of coherence measures. The convex
roof construction is as follows. We first choose a concave function $%
f(p_{1},p_{2},...,p_{d})$ defined on the probability distribution space $%
(p_{1},p_{2},...,p_{d})$, $f(p_{1},p_{2},...,p_{d})$ is invariant under the
index permutation of $\{j\}_{j=1}^{d},$ $f(p_{1},p_{2},...,p_{d})\geq 0$ and
$f(1,0,0,...,0)=0.$ The coherence of pure state $|\psi \rangle $ is defined
as
\begin{equation}
C_{f}(|\psi \rangle )=f(|\langle 1|\psi \rangle |^{2},|\langle 2|\psi
\rangle |^{2},|\langle 3|\psi \rangle |^{2},...,|\langle d|\psi \rangle
|^{2}),     \label{eq2B-8}
\end{equation}
the coherence of mixed state $\rho $ is defined as
\begin{equation}
C_{f}(\rho )=\min_{\{q_{\mu },|\varphi _{\mu }\rangle \}_{\mu }}\sum_{\mu
}q_{\mu }C_{f}(|\varphi _{\mu }\rangle ),     \label{eq2B-9}
\end{equation}%
where min runs over all pure state decompositions $\rho =\sum_{\mu }q_{\mu
}|\varphi _{\mu }\rangle \langle \varphi _{\mu }|.$

We have reviewed some existing coherence measures. With these coherence measures, we have Theorem \ref{Theorem-3} below.

\begin{Theorem} \label{Theorem-3}
Any coherence measure $C$ defined in Eqs. (\ref{eq2B-1}-\ref{eq2B-9}) satisfies $%
C(\rho )=C(\rho ^{\ast }).$
\end{Theorem}

\emph{Proof.} One can check that coherence measures defined in Eqs. (\ref{eq2B-1}-\ref{eq2B-7}) certainly
satisfy $C(\rho )=C(\rho ^{\ast }).$ In Eq. (\ref{eq2B-5}), notice that $F(\rho ,\sigma
)=F(\rho ^{\ast },\sigma ^{\ast })$ for any states $\rho $ and $\sigma ,$
then $C_{\text{g}}(\rho )=C_{\text{g}}(\rho ^{\ast }).$

We then only consider the convex roof coherence measures. For a convex roof
coherence measure $C_{f}(\rho )$ defined in Eqs. (\ref{eq2B-8},\ref{eq2B-9}), Eq. (\ref{eq2B-8}) implies $%
C_{f}(|\psi \rangle )=C_{f}(|\psi ^{\ast }\rangle )$, with the fact that if $%
\rho =\sum_{\mu }q_{\mu }|\varphi _{\mu }\rangle \langle \varphi _{\mu }|$
is a pure state decomposition of $\rho $ then $\rho ^{\ast }=\sum_{\mu
}q_{\mu }|\varphi _{\mu }^{\ast }\rangle \langle \varphi _{\mu }^{\ast }|$
is a pure state decomposition of $\rho ^{\ast },$ Eq. (\ref{eq2B-9}) thus implies $%
C_{f}(\rho )=C_{f}(\rho ^{\ast }).$
$\hfill\square$

Theorem \ref{Theorem-3} and Theorem \ref{Theorem-1} together imply that coherence measures defined in
Eqs. (\ref{eq2B-1}-\ref{eq2B-9}) all satisfy $C(\rho )-C($Re$\rho )\geq 0.$

\subsection{Examples}
Let $U=\sum_{j=1}^{d}e^{i\theta _{j}}|j\rangle \langle j|$ with $\{\theta
_{j}\}_{j=1}^{d}$ real numbers, i.e., $U$ is a diagonal unitary. Since $%
U\sigma U^{\dagger }$ is diagonal for any diagonal state $\sigma ,$ then $U$
is an incoherent channel and (C2) implies $C(U\rho U^{\dagger })\leq C(\rho )
$ for any state $\rho .$ Notice that $U^{\dagger }=\sum_{j=1}^{d}e^{-i\theta
_{j}}|j\rangle \langle j|$ is also a diagonal unitary and an incoherent
channel, thus $C(\rho )=C[U^{\dagger }(U\rho U^{\dagger })U]\leq C(U\rho
U^{\dagger }).$ As a result,
\begin{equation}
C(\rho )=C(U\rho U^{\dagger })    \label{eq2C-1}
\end{equation}%
for any any state $\rho $ and any diagonal unitary $U.$ Eq. (\ref{eq2C-1}) will be
useful in Example 1 and Example 2.

\begin{Example} \label{Example-1}
For any coherence measure $C$ and any pure state $|\psi
\rangle $, it holds that $C(|\psi \rangle )=C(|\psi ^{\ast }\rangle ).$
\end{Example}

\emph{Proof.} Expand $|\psi \rangle $ in basis $\{|j\rangle \}_{j=1}^{d}$ as
$|\psi \rangle =\sum_{j=1}^{d}|\langle j|\psi \rangle |e^{i\theta
_{j}}|j\rangle $ with $\{\theta _{j}\}_{j=1}^{d}$ real numbers. Let $%
U=\sum_{j=1}^{d}e^{-2i\theta _{j}}|j\rangle \langle j|,$ then $U|\psi
\rangle =|\psi ^{\ast }\rangle ,$ and Eq. (\ref{eq2C-1}) yields $C(|\psi \rangle
)=C(|\psi ^{\ast }\rangle ).$
$\hfill\square$

\begin{Example} \label{Example-2}
For any coherence measure $C$ and any qubit state $\rho $,
it holds that $C(\rho )=C(\rho ^{\ast }).$
\end{Example}

\emph{Proof.} Expand $\rho $ in basis $\{|j\rangle \}_{j=1}^{2}$ as $\rho
=\sum_{j,k=1}^{2}\rho _{jk}|j\rangle \langle k|.$ For clarity we write $\rho
=\sum_{j,k=1}^{2}\rho _{jk}|j\rangle \langle k|$ in the matrix form
\begin{equation*}
\rho =\left(
\begin{array}{cc}
\rho _{11} & |\rho _{12}|e^{i\theta } \\
|\rho _{12}|e^{-i\theta } & \rho _{22}%
\end{array}%
\right)
\end{equation*}
with $\theta $ a real number. Let $U=$diag$(1,e^{2i\theta }),$ then $U\rho
U^{\dagger }=\rho ^{\ast }$ and Eq. (\ref{eq2C-1}) yileds $C(\rho )=C(\rho ^{\ast }).$
$\hfill\square$

\begin{figure}[!htb]
\includegraphics[width=8cm]{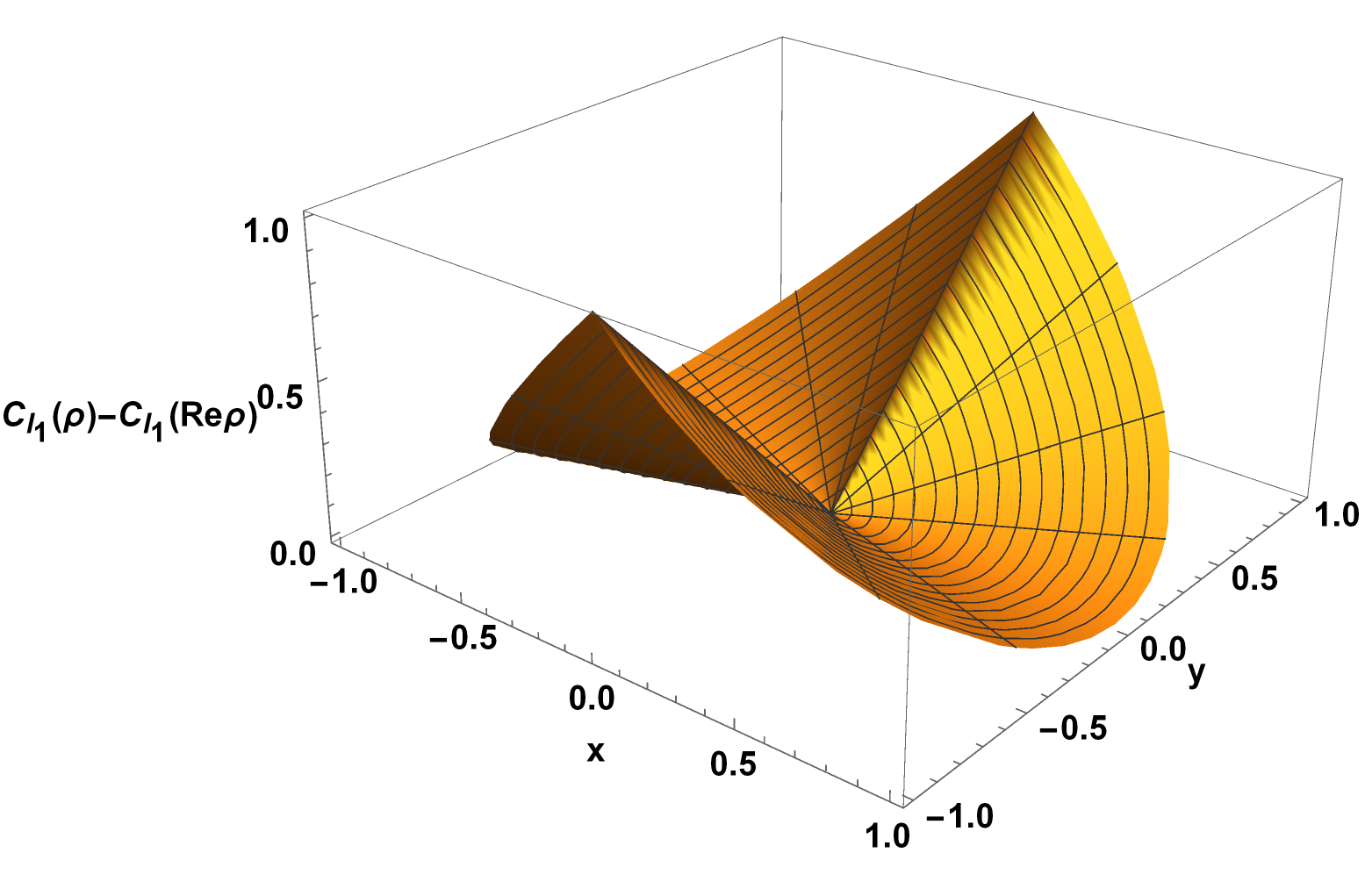}
\caption{$C_{l_{1}}(\rho )-C_{l_{1}}(\text{Re}\rho )$  versus $x$ and $y$ in Eq. (\ref{eq2C-4}).}
\label{Fig1}
\end{figure}

\begin{Example} \label{Example-3}
For the coherence measure $C_{l_{1}}$ defined in Eq. (\ref{eq2B-1}), one has
\begin{equation}
C_{l_{1}}(\rho )-C_{l_{1}}(\text{Re}\rho )=\sum_{j,k=1}^{d}(|\rho _{jk}|-|%
\text{Re}\rho _{jk}|).    \label{eq2C-2}
\end{equation}
\end{Example}
We see that $C_{l_{1}}(\rho )-C_{l_{1}}(\text{Re}\rho )\geq 0$ and $C_{l_{1}}(\rho )-C_{l_{1}}(\text{Re}\rho )= 0$ if and only if $\rho$ is real. 
When $d=2,$ we express state $\rho =\sum_{j,k=1}^{2}\rho _{jk}|j\rangle
\langle k|$ in the Bloch representation as
\begin{equation}
\rho =\frac{1}{2}\left(
\begin{array}{cc}
1+z & x-iy \\
x+iy & 1-z%
\end{array}%
\right),     \label{eq2C-3}
\end{equation}
with $\{x,y,z\}$ real numbers satisfying $x^{2}+y^{2}+z^{2}\leq 1.$ For this
case, Eq. (\ref{eq2C-2}) reads
\begin{equation}
C_{l_{1}}(\rho )-C_{l_{1}}(\text{Re}\rho )=\sqrt{x^{2}+y^{2}}-|x|.    \label{eq2C-4}
\end{equation}
We depict Eq. (\ref{eq2C-4}) in Fig. \ref{Fig1}.

\section{Coherence and imaginarity of Gaussian states}   \label{Section-III}

\subsection{Background of (bosonic) Gaussian states}
Gaussian states in bosonic systems are widely used in quantum physics such as quantum optics and quantum information science \cite{RMP-2005-Braunstein,PR-2007-Wang,arXiv-2005-Ferraro,EPJ-2012-Olivares,
RMP-2012-Weedbrook,OSID-2014-Adesso,book-2017-Serafini}. Coherence theory and imaginarity theory are all dependent on an orthonormal basis. Fock basis is the orthonormal basis of the Hilbert space for Gaussian states, then it is natural to establish the coherence theory and imaginarity theory of Gaussian states with respect to Fock basis. However, by the definition of Gaussian states in Eq. (\ref{eq3A-11}), it is convenient to express Gaussian states by the means and covariance matrices, but difficult to express general Gaussian states in Fock basis (some explorations about expressing general Gaussian states in Fock basis see for examples \cite{PRA-Xu-2016,PRA-2019-Quesada,JPA-2020-Huh,arxiv-2022-Yao,PRA-2023-Xu}). Therefore, for coherence theory and imaginarity theory of Gaussian states, a great challenge is to find coherence measures or imaginarity measures which can be expressed by the means and covariance matrices.

We first review some basics of Gaussian states and introduce the notation we will use. Let $\{|j\rangle \}_{j=0}^{\infty }$ be an orthonormal basis with $j\in
\{0,1,2,...\}.$ $\{|j\rangle \}_{j=0}^{\infty }$ spans the complex Hilbert
space $H$ over complex numbers. $\{|j\rangle \}_{j=0}^{\infty }$ is called
the one-mode Fock basis. The $n$-mode Fock basis is $\{|j\rangle
\}_{j}^{\otimes n},$ the $n$-fold tensor product of $\{|j\rangle
\}_{j=0}^{\infty },$ and $\{|j\rangle \}_{j}^{\otimes n}$ spans the complex
Hilbert space $\otimes _{l=1}^{n}H_{l}=H^{\otimes n}$ over complex numbers
with each $H_{l}=H.$ Below we discuss the coherence and imaginarity of
Gaussian states under the Fock basis $\{|j\rangle \}_{j}^{\otimes n}.$

On each $H_{l},$ the annihilation operator and creation operator are defined as
\begin{eqnarray}
\widehat{a}_{l}|0\rangle &=&0,\ \ \widehat{a}_{l}|j\rangle =\sqrt{j}%
|j-1\rangle \text{ for }j\geq 1;  \label{eq3A-1} \\
\widehat{a}_{l}^{\dagger }|j\rangle &=&\sqrt{j+1}|j+1\rangle \text{ for }%
j\geq 0.  \label{eq3A-2}
\end{eqnarray}
From $\{\widehat{a}_{l},\widehat{a}_{l}^{\dagger }\}_{l=1}^{n}$ we define $\{%
\widehat{q}_{l},\widehat{p}_{l}^{\dagger }\}_{l=1}^{n}$ as
\begin{equation}
\widehat{q}_{l}=\widehat{a}_{l}+\widehat{a}_{l}^{\dagger },\ \ \widehat{p}%
_{l}=-i(\widehat{a}_{l}-\widehat{a}_{l}^{\dagger }),  \label{eq3A-3}
\end{equation}%
We arrange $\{\widehat{q}_{l},\widehat{p}_{l}^{\dagger }\}_{l=1}^{n}$ as a
vector as
\begin{eqnarray}
\widehat{X} &=&(\widehat{q}_{1},\widehat{p}_{1},\widehat{q}_{2},\widehat{p}%
_{2},...,\widehat{q}_{n},\widehat{p}_{n})^{T}  \notag \\
&=&(\widehat{X}_{1},\widehat{X}_{2},\widehat{X}_{3},\widehat{X}_{4},...,%
\widehat{X}_{2n-1},\widehat{X}_{2n})^{T}.  \label{eq3A-4}
\end{eqnarray}

A quantum state $\rho $ in $\overline{H}^{\otimes n}$ can be characterized
by its characteristic function
\begin{equation}
\chi (\rho ,\xi )=\text{tr}[\rho D(\xi )],  \label{eq3A-5}
\end{equation}%
where $D(\xi )$ is the displacement operator
\begin{eqnarray}
D(\xi ) &=&\exp (i\widehat{X}^{T}\Omega \xi ),   \label{eq3A-6}  \\
\Omega  &=&\oplus _{l=1}^{n}\omega ,\ \ \omega =\left(
\begin{array}{cc}
0 & 1 \\
-1 & 0%
\end{array}%
\right) ,  \label{eq3A-7}   \\
\xi  &=&(\xi _{1},\xi _{2},...,\xi _{2n})^{T}\in
\mathbb{R}^{2n}.   \notag
\end{eqnarray}

For state $\rho $ in $\overline{H}^{\otimes n},$ the mean of $\rho $ is
\begin{equation}
\overline{X}=\text{tr}(\rho \widehat{X})=(\overline{X}_{1},\overline{X}%
_{2},...,\overline{X}_{2n})^{T};   \label{eq3A-8}
\end{equation}%
the covariance matrix $V$ is defined by its elements
\begin{equation}
V_{lm}=\frac{1}{2}\text{tr}(\rho \{\Delta \widehat{X}_{l},\Delta \widehat{X}%
_{m}\})   \label{eq3A-9}
\end{equation}%
where $\Delta \widehat{X}_{l}=\widehat{X}_{l}-\overline{X}_{l}$, and $%
\{\Delta \widehat{X}_{l},\Delta \widehat{X}_{m}\}=\Delta \widehat{X}%
_{l}\Delta \widehat{X}_{m}+\Delta \widehat{X}_{m}\Delta \widehat{X}_{l}$ is
the anticommutator of $\Delta \widehat{X}_{l}$ and $\Delta \widehat{X}_{m}.$
The covariance matrix $V=V^{T}$ is a $2n\times 2n$ real and symmetric matrix
satisfying the uncertainty principle \cite{PRA-1994-Simon}
\begin{equation}
V+i\Omega \geq 0.  \label{eq3A-10}
\end{equation}%
Note that $V+i\Omega \geq 0$ implies $V>0$ meaning that $V$ is positive
definite.

A quantum state $\rho $ in $\overline{H}^{\otimes n}$ is called an $n$-mode
Gaussian state if its characteristic function has the Gaussian form
\begin{equation}
\chi (\rho ,\xi )=\exp \left[-\frac{1}{2}\xi ^{T}(\Omega V\Omega ^{T})\xi
-i(\Omega \overline{X})^{T}\xi \right],   \label{eq3A-11}
\end{equation}%
where $\overline{X}$ is the mean of $\rho $ and $V$ is the covariance matrix
of $\rho .$ $\overline{X}$ and $V$ with Eq. (\ref{eq3A-10}) completely
determine the Gaussian state $\rho $ \cite{PRA-1994-Simon}, thus we write $%
\rho $ as $\rho (\overline{X},V).$

A Gaussian channel $\phi $ on $\overline{H}^{\otimes n}$ can be represented
by $\phi =(b,T,N)$, here $b=(b_{1},b_{2},...,b_{2n})^{T}\in\mathbb{R}^{2n},$ $T$ and $N=N^{T}$ are $2n\times 2n$ real matrices. $\phi =(b,T,N)$
maps the Gaussian state $\rho (\overline{X},V)$ to the Gaussian state with
mean and covariance matrix
\begin{equation}
\overline{X}\rightarrow T\overline{X}+b,\ \ V\rightarrow TVT^{T}+N,   \label{eq3A-12}
\end{equation}%
and $\phi =(b,T,N)$ fulfils the complete positivity condition
\begin{equation}
N+i\Omega -iT\Omega T^{T}\geq 0.  \label{eq3A-13}
\end{equation}

After introducing the notation of Gaussian states and Gaussian channels, we
turn to the imaginarity and coherence of Gaussian states.

The (complex) conjugate of the Gaussian state $\rho (\overline{X},V)$ is
still a Gaussian state $\rho ^{\ast }(O\overline{X},OVO)$ with \cite{PRA-2023-Xu}
\begin{equation}
O=\oplus _{l=1}^{n}\left(
\begin{array}{cc}
1 & 0 \\
0 & -1%
\end{array}%
\right) .            \label{eq3A-14}
\end{equation}
As a result, a Gaussian state $\rho (\overline{X},V)$ is real if and only if $%
\rho (\overline{X},V)=\rho ^{\ast }(O\overline{X},OVO),$ that is, $\overline{%
X}=O\overline{X}$ and $V=OVO.$ From Gaussian state $\rho (\overline{X},V),$
we can define the real Gaussian state $\rho ^{\prime }\left(\frac{\overline{X}+O%
\overline{X}}{2},\frac{V+OVO}{2}\right).$ Obviously, for Gaussian state $\rho (%
\overline{X},V),$ it holds that
\begin{equation}
\rho ^{\prime }=(\rho ^{\prime })^{\ast }=({\rho ^{\ast }})^{\prime },  \label{eq3A-15}
\end{equation}%
and $\rho $ is real (i.e., $\rho =$Re$\rho $) if and only if $\rho =\rho
^{\prime }.$ In general, Re$\rho \neq \rho ^{\prime }$ for Gaussian state $%
\rho (\overline{X},V).$

It is shown \cite{PRA-Xu-2016} that a Gaussian state $\rho (%
\overline{X},V)$ is diagonal if and only if $\rho (\overline{X},V)$ is a
thermal state, that is, $\rho (\overline{X},V)=\rho (0,\oplus _{l=1}^{n}\nu
_{j}I_{2})$ with $\nu _{j}\geq 1$ for all $j.$ A Gaussian channel $\phi $ is
called incoherent if $\phi (\sigma )$ is still diagonal for any thermal
state $\sigma .$

For Gaussian states $\rho $ and $\sigma $, the convex combination $p\rho
+(1-p)\sigma $ in general is not Gaussian, here $p\in (0,1).$ This fact
makes troubles to adopt (C3) and (C4) as necessary conditions for coherence
measures of Gaussian states. With this consideration, we abandon (C3) and (C4), and modify (C1) and (C2) for Gaussian states as (CG1) and (CG2) below.
A real valued functional $C_{\text{G}}$ defined on Gaussian states is called a
coherence measure if $C_{\text{G}}$ satisfies the following (CG1) and (CG2).

\textbf{(CG1)} Faithfulness. $C_{\text{G}}(\rho )\geq 0$ for any Gaussian state $\rho ,$
and $C_{\text{G}}(\rho )=0$ if and only if $\rho $ is diagonal.

\textbf{(CG2)} Monotonicity. $C_{\text{G}}[\phi (\rho )]\leq C_{G}(\rho )$ for any
incoherent Gaussian channel $\phi .$

Coherence measures, incoherent channels and state transformations for Gaussian states were investigated recently \cite{arXiv-Illuminati-2016,PLA-Xu-2021,PRA-Du-2022,PRA-Du-2023}.

\subsection{Coherence and imaginarity of Gaussian states}

To study the relationship between coherence and imaginarity of Gaussian
states, we propose the conditions (CG4) and (CG6) below, (CG6) can be viewed
as the counterpart of (C6).

\textbf{(CG4)} $C_{\text{G}}(\rho ^{\prime })\leq \frac{1}{2}[C_{\text{G}}(\rho )+C_{\text{G}}(\rho ^{\ast
})]$ for any Gaussian state $\rho .$

\textbf{(CG6)} $C_{\text{G}}(\rho )=C_{\text{G}}(\rho ^{\ast })$ for any Gaussian state $\rho .$

It is easy to see that if a coherence measure $C_{\text{G}}(\rho )$ for Gaussian
states satisfies (CG4) and (CG6), then $C_{\text{G}}(\rho )-C_{\text{G}}(\rho ^{\prime
})\geq 0.$

Similar to definition \ref{Definition-1}, we give the definition of the conjugate of a
Gaussian channel.
\begin{Definition}   \label{Definition-2}
For the Gaussian channel $\phi =(b,T,N),$ we define the
conjugate of $\phi =(b,T,N)$ as $\phi ^{\ast }=(Ob,OTO,ONO).$
\end{Definition}

From Eqs. (\ref{eq3A-13},\ref{eq3A-14}), we have
\begin{equation*}
O(N+i\Omega -iT\Omega T^{T})O\geq 0.
\end{equation*}
Using the facts $O\Omega =-\Omega O,$ $O^{T}=O$ and $O^{2}=I_{2n},$ we obtain
\begin{equation*}
ONO-i\Omega +i(OTO)\Omega (OTO)^{T}\geq 0.
\end{equation*}
Taking the conjugate of the left side of above equation and using the facts $N^{T}=N$ and $\Omega^{T}=-\Omega$ yield
\begin{equation*}
ONO+i\Omega -i(OTO)\Omega (OTO)^{T}\geq 0.
\end{equation*}
This show that $\phi ^{\ast }=(Ob,OTO,ONO)$ satisfies the complete
positivity condition, then $\phi ^{\ast }$ is a Gaussian channel.

With the definition of $\phi ^{\ast }=(Ob,OTO,ONO)$ and Eq. (\ref{eq3A-12}), one can
check that
\begin{equation}
\phi ^{\ast }(\rho ^{\ast })=[\phi (\rho )]^{\ast }.   \label{eq3B-1}
\end{equation}

When a Gaussian channel $\phi =(d,T,N)$ is incoherent, we have Corollary 1 as follows.

\begin{Corollary}   \label{Corollary-1}
If the Gaussian channel $\phi =(b,T,N)$ is incoherent,
then its conjugate $\phi ^{\ast }=(Ob,OTO,ONO)$ is also incoherent, and $%
\phi ^{\ast }(\sigma )=\phi (\sigma )$ for any thermal state $\sigma .$
\end{Corollary}

\emph{Proof.} For any thermal state $\sigma ,$ we have $\sigma =\sigma
^{\ast }.$ Since $\phi $ is incoherent, then $\phi (\sigma )$ is thermal.
From Eq. (\ref{eq3B-1}) one has $\phi ^{\ast }(\sigma )=\phi ^{\ast }(\sigma ^{\ast
})=[\phi (\sigma )]^{\ast }=\phi (\sigma ).$
$\hfill\square$

For $p\in(0,1)$, Gaussian states $\rho$ and $\sigma$, the convex combination of $\rho$ and $\sigma$, $p\rho+(1-p)\sigma$ is not Gaussian in general. Here we define a Gaussian state by the convex combinations of the means and covariance matrices of $\rho$ and $\sigma$.

\begin{Definition} \label{Definition-3}
For $p\in (0,1)$, Gaussian states $\rho (\overline{X}%
,V) $ and $\sigma (\overline{Y},W),$ we define the Gaussian state $\tau (p%
\overline{X}+(1-p)\overline{Y},pV+(1-p)W),$ and denote
\begin{eqnarray}
&&p\rho (\overline{X},V)\boxplus (1-p)\sigma (\overline{Y},W) \notag \\
&=&\tau (p\overline{X}+(1-p)\overline{Y},pV+(1-p)W).    \label{eq3B-2}
\end{eqnarray}
\end{Definition}

Under Definition \ref{Definition-3}, we see that
\begin{eqnarray}
&&\rho ^{\prime }\left(\frac{\overline{X}+O\overline{X}}{2},\frac{V+OVO}{2}\right)  \notag \\
&=&\frac{1}{2}\rho (\overline{X},V)\boxplus \frac{1}{2}\rho ^{\ast }(O\overline{X}%
,OVO).     \label{eq3B-3}
\end{eqnarray}

Since Gaussian states $\rho (\overline{X},V)$ and $\sigma (\overline{Y},W)
$ satisfy the uncertainty principle $V+i\Omega \geq 0$ and $W+i\Omega \geq 0,
$ then
\begin{eqnarray*}
&&\lbrack pV+(1-p)W]+i\Omega  \\
&=&p(V+i\Omega )+(1-p)(W+i\Omega )\geq 0.
\end{eqnarray*}
This says $pV+(1-p)W$ satisfies the uncertainty principle and $\tau (p%
\overline{X}+(1-p)\overline{Y},pV+(1-p)W)$ indeed is a Gaussian state.

Similar to Theorem \ref{Theorem-2}, we have Theorem \ref{Theorem-4} below.

\begin{Theorem}    \label{Theorem-4}
If $C_{\text{G}}(\rho )$ is a coherence measure for Gaussian
states, then
\begin{equation}
C_{\text{G}}^{\prime }(\rho )=\frac{1}{2}[C_{\text{G}}(\rho )+C_{\text{G}}(\rho ^{\ast })]    \label{eq3B-4}
\end{equation}
is also a coherence measure for Gaussian states, and $C_{\text{G}}^{\prime }(\rho )$
satisfies $C_{\text{G}}^{\prime }(\rho )=C_{\text{G}}^{\prime }(\rho ^{\ast }).$
\end{Theorem}

\emph{Proof.} $C_{\text{G}}^{\prime }(\rho )$ satisfying (CG1) is easy to prove. We
now prove that $C_{\text{G}}^{\prime }(\rho )$ satisfies (CG2). For any Gaussian
state $\rho $ and any incoherent Gaussian channel $\phi,$ we have
\begin{eqnarray*}
C_{\text{G}}^{\prime }[\phi (\rho )] &=&\frac{1}{2}[C_{\text{G}}(\phi (\rho ))+C_{\text{G}}([\phi
(\rho )]^{\ast })] \\
&=&\frac{1}{2}[C_{\text{G}}(\phi (\rho ))+C_{\text{G}}(\phi ^{\ast }(\rho ^{\ast }))] \\
&\leq &\frac{1}{2}[C_{\text{G}}(\rho )+C_{\text{G}}(\rho ^{\ast })]=C_{\text{G}}^{\prime }(\rho ),
\end{eqnarray*}
where we have used $\phi ^{\ast }(\rho ^{\ast })=[\phi (\rho )]^{\ast }$,
the fact that $\phi ^{\ast }$ is incoherent, and (CG2) that $C_{\text{G}}(\phi
(\rho ))\leq C_{\text{G}}(\rho )$ and $C_{\text{G}}(\phi ^{\ast }(\rho ^{\ast }))\leq
C_{\text{G}}(\rho ^{\ast }).$
$\hfill\square$

In Ref. \cite{PRA-Xu-2016}, the author introduced the Gaussian relative entropy $C_{\text{Gr}}(\rho )$ for
Gaussian state $\rho$ as
\begin{eqnarray}
C_{\text{Gr}}(\rho ) &=&S(\rho ||\overline{\rho })=S(\overline{\rho }%
)-S(\rho ),   \label{eq3B-5}   \\
\overline{\nu _{j}} &=&\frac{1}{2}(V_{2j-1,2j-1}+V_{2j,2j}+\overline{X}%
_{2j-1}^{2}+\overline{X}_{2j}^{2}),    \ \ \ \     \label{eq3B-6}
\end{eqnarray}
where $\rho =\rho (\overline{X},V)$ is any Gaussian state, $\overline{\rho }$
$=\overline{\rho }(0,\oplus _{j=1}^{n}\overline{\nu _{j}}I_{2})$ is a
thermal state induced by $\rho (\overline{X},V),$ $S(\rho )=-$tr$(\rho \log
_{2}\rho )$ is the von Neumann entropy of $\rho $, $S(\rho ||\overline{\rho }%
)=$tr$(\rho \log _{2}\rho )-$tr$(\rho \log _{2}\overline{\rho })$ is the
relative entropy of $\rho $ to $\overline{\rho }.$ The defining property of the thermal state $\overline{\rho }(0,\oplus _{j=1}^{n}\overline{\nu _{j}}I_{2})$ is $\text{tr}(\widehat{a}_{l}^{\dagger }\widehat{a}_{l}\rho)=\text{tr}(\widehat{a}_{l}^{\dagger }\widehat{a}_{l}\overline{\rho })$ for all $\{j\}_{j=1}^{n}$, namely, $\rho$ and $\overline{\rho }$ have the same mean particle numbers for all modes.

$S(\overline{\rho })$ and
$S(\rho )$ in Eq. (\ref{eq3B-5}) can be computed by \cite{PRA-Holevo-1999}
\begin{eqnarray}
S(\rho ) &=&\sum_{j=1}^{n}g(\nu _{j}),   \label{eq3B-7}  \\
S(\overline{\rho }) &=&\sum_{j=1}^{n}g(\overline{\nu _{j}}),   \label{eq3B-8}  \\
g(x) &=&\frac{x+1}{2}\log _{2}\frac{x+1}{2}-\frac{x-1}{2}\log _{2}\frac{x-1}{%
2}, \label{eq3B-9}
\end{eqnarray}
where $\{\nu _{j}\}_{j=1}^{n}$ is the symplectic eigenvalues of $V$ (see
\hyperlink{Appendix}{Appendix} for more details).

It is shown that \cite{PRA-Xu-2016} $C_{\text{Gr%
}}(\rho )$ satisfies (CG1) and (CG2), i.e., $C_{\text{Gr}}(\rho )$ is a
coherence measure for Gaussian states. Now we show that $C_{\text{Gr}}(\rho )
$ also satisfies (CG4) and (CG6).

$C_{\text{Gr}}(\rho )$ satisfying (CG6) is easy to prove. By the definition $%
S(\rho )=-$tr$(\rho \log _{2}\rho )$ we see that $S(\rho )=S(\rho ^{\ast }).$
Alternatively, $S(\rho )=S(\rho ^{\ast })$ follows by Eq. (\ref{eq3B-7}) and the fact
that the Gaussian state $\rho (\overline{X},V)$ and its conjugate $\rho (O%
\overline{X},OVO)$ have the same symplectic eigenvalues (see \hyperlink{Appendix}{Appendix} for
more details). $\overline{\rho }=\overline{\rho ^{\ast }}$ evidently holds
from Eq. (\ref{eq3B-6}). Then Eq. (\ref{eq3B-5}) ensures that $C_{\text{Gr}}(\rho )$ satisfies (CG6).

Since $C_{\text{Gr}}(\rho )$ satisfies (CG6), then $C_{\text{Gr}}(\rho )$
satisfying (CG4) is equivalent to the following Theorem \ref{Theorem-5}.

\begin{Theorem} \label{Theorem-5}
$C_{\text{Gr}}(\rho )-C_{\text{Gr}}(\rho ^{\prime })\geq 0$
for any Gaussian state $\rho .$
\end{Theorem}

\emph{Proof. }From Eqs. (\ref{eq3B-5}-\ref{eq3B-9}) we have
\begin{eqnarray}
&&C_{\text{Gr}}(\rho )-C_{\text{Gr}}(\rho ^{\prime }) \notag \\
&=&S(\overline{\rho })-S(\rho )-S(\overline{\rho ^{\prime }})+S(\rho
^{\prime })  \notag  \\
&=&[S(\overline{\rho })-S(\overline{\rho ^{\prime }})]+[S(\rho ^{\prime
})-S(\rho )],  \label{eq3B-10}  \\
S(\overline{\rho }) &=&\sum_{j=1}^{n}g\left[\frac{1}{2}(V_{2j-1,2j-1}+V_{2j,2j}+%
\overline{X}_{2j-1}^{2}+\overline{X}_{2j}^{2})\right], \notag \\ \label{eq3B-11}  \\
S(\overline{\rho ^{\prime }}) &=&\sum_{j=1}^{n}g\left[\frac{1}{2}%
(V_{2j-1,2j-1}+V_{2j,2j}+\overline{X}_{2j-1}^{2})\right].  \label{eq3B-12}
\end{eqnarray}
Since $g(x)$ is increasing, then $S(\overline{\rho })-S(\overline{\rho
^{\prime }})\geq 0.$ Corollary \ref{Corollary-2} in Appendix ensures $S(\rho ^{\prime
})-S(\rho )\geq 0.$ Theorem \ref{Theorem-5} then follows.
$\hfill\square$

\subsection{Examples}
\begin{figure}[!htb]
\includegraphics[width=8cm]{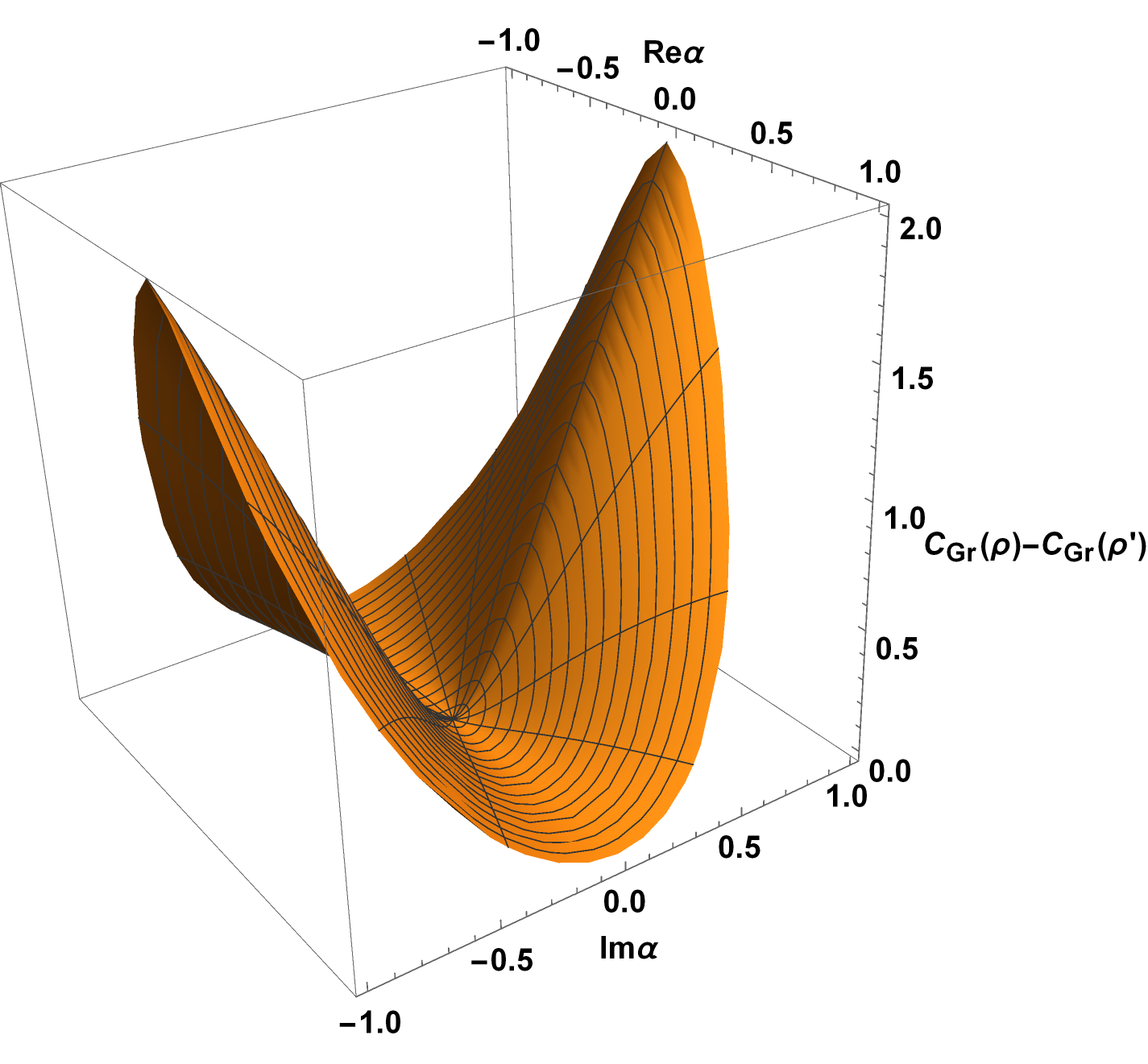}
\caption{$C_{\text{Gr}}(\rho )-C_{\text{Gr}}(\rho ^{\prime })$  versus $\text{Re}\alpha$ and $\text{Im}\alpha$ in Eq. (\ref{eq3C-2}).}
\label{Fig2}
\end{figure}

\begin{Example} \label{Example-4}
Consider the one-mode Glauber coherent state
\begin{equation}
|\alpha \rangle =e^{-\frac{|\alpha |^{2}}{2}}\sum_{j=0}^{\infty }\frac{%
\alpha ^{j}}{\sqrt{j!}}|j\rangle    \label{eq3C-1}
\end{equation}%
\end{Example}
with $\alpha $ any complex number. The mean of $|\alpha \rangle \langle
\alpha |$ is $\overline{X}=(2\text{Re}\alpha ,2\text{Im}\alpha )^{T},$ the
covariance matrix of $\rho =|\alpha \rangle \langle \alpha |$ is $V=\left(
\begin{array}{cc}
1 & 0 \\
0 & 1%
\end{array}%
\right) .$ Then $\rho ^{\prime }\left(\frac{\overline{X}+O\overline{X}}{2},\frac{%
V+OVO}{2}\right)$ has $\frac{\overline{X}+O\overline{X}}{2}=(2$Re$\alpha ,0)^{T},%
\frac{V+OVO}{2}=V=\left(
\begin{array}{cc}
1 & 0 \\
0 & 1%
\end{array}%
\right) .$ Eqs. (\ref{eq3B-10},\ref{eq3B-11},\ref{eq3B-12}) yield $%
S(\rho ^{\prime })-S(\rho )=0$ and
\begin{eqnarray}
&&C_{\text{Gr}}(\rho )-C_{\text{Gr}}(\rho ^{\prime })=S(\overline{\rho })-S(%
\overline{\rho ^{\prime }}) \notag  \\
&=&g[1+2(\text{Re}\alpha )^{2}+2(\text{Im}\alpha )^{2}]-g[1+2(\text{Re}%
\alpha )^{2}].   \label{eq3C-2}
\end{eqnarray}
We see that, $C_{\text{Gr}}(\rho )-C_{\text{Gr}}(\rho ^{\prime })\geq 0$ and $C_{\text{Gr}}(\rho )-C_{\text{Gr}}(\rho ^{\prime })=0$ if and only if $\alpha$ is a real number, namely, $|\alpha \rangle \langle
\alpha |$ is a real Gaussian state. We depict Eq. (\ref{eq3C-2}) in Fig. \ref{Fig2}.

\begin{Example} \label{Example-5}
Consider the one-mode squeezed state
\begin{eqnarray}
|\zeta \rangle  &=&\exp \left[\frac{1}{2}(\zeta ^{\ast }\widehat{a}^{2}-\zeta
\widehat{a}^{\dagger 2})\right]|0\rangle   \notag \\
&=&\frac{1}{\sqrt{\cosh |\zeta |}}\sum_{j=0}^{\infty }(-e^{i\theta }\tanh
|\zeta |)^{j}\frac{\sqrt{(2j)!}}{2^{j}j!}|2j\rangle ,\ \    \label{eq3C-3}
\end{eqnarray}%
\end{Example}
with $\zeta $ any complex number and $\zeta =|\zeta |e^{i\theta }$ its polar
form. $\exp [\frac{1}{2}(\zeta ^{\ast }\widehat{a}^{2}-\zeta \widehat{a}%
^{\dagger 2})]$ is the squeezing operator. The mean of $|\zeta \rangle
\langle \zeta |$ is $\overline{X}=(0,0)^{T},$ the covariance matrix $V$ of $%
|\zeta \rangle \langle \zeta |$ is
\begin{equation}
\begin{cases}
V_{11}=\cosh (2|\zeta |)+\cos \theta \sinh (2|\zeta |) \\
V_{12}=V_{21}=\sin \theta \sinh (2|\zeta |) \\
V_{22}=\cosh (2|\zeta |)-\cos \theta \sinh (2|\zeta |).%
\end{cases}   \label{eq3C-4}
\end{equation}%
Then Eqs. (\ref{eq3B-10},\ref{eq3B-11},\ref{eq3B-12}) yield $S(%
\overline{\rho })-S(\overline{\rho ^{\prime }})=0$. With the facts $S(\rho )=\text{det}V$ and $S(\rho ^{\prime
})=\text{det}\frac{V+OVO}{2}$ (see more details in \hyperlink{Appendix}{Appendix}), one gets
\begin{eqnarray}
&&C_{\text{Gr}}(\rho )-C_{\text{Gr}}(\rho ^{\prime })=S(\rho ^{\prime
})-S(\rho )   \notag  \\
&=&g[1+\sin ^{2}\theta \sinh ^{2}(2|\zeta |)].   \label{eq3C-5}
\end{eqnarray}%

\begin{figure}[!htb]
\includegraphics[width=8cm]{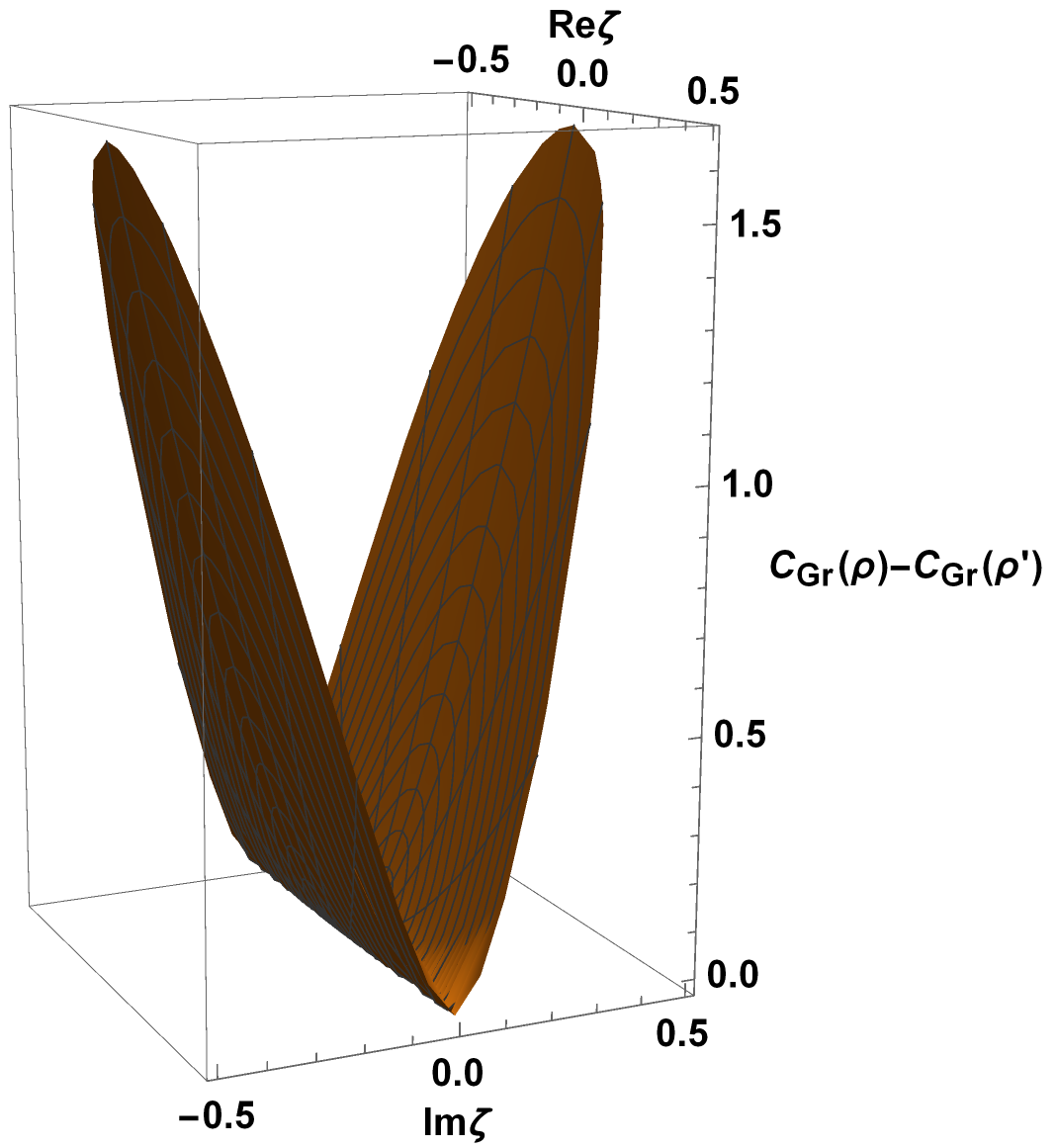}
\caption{$C_{\text{Gr}}(\rho )-C_{\text{Gr}}(\rho ^{\prime })$  versus $\text{Re}\zeta$ and $\text{Im}\zeta$ in Eq. (\ref{eq3C-5}).}
\label{Fig3}
\end{figure}

We see that, $C_{\text{Gr}}(\rho )-C_{\text{Gr}}(\rho ^{\prime })\geq 0$ and $C_{\text{Gr}}(\rho )-C_{\text{Gr}}(\rho ^{\prime })=0$ if and only if $\zeta$ is a real number, that is, $|\zeta \rangle
\langle \zeta |$ is a real Gaussian state. We depict Eq. (\ref{eq3C-5}) in Figure \ref{Fig3}.

\section{Summary}   \label{Section-IV}
For the fixed orthonormal basis, if a quantum state $\rho $ has nonzero
imaginary part then $\rho $ must be coherent. We discussed how to
quantitatively characterize this fact. We proved that, if a coherence
measure $C$ satisfies $C(\rho )=C(\rho ^{\ast })$ then $C(\rho )-C($Re$\rho
)\geq 0.$ If a coherence measure $C$ does not satisfy $C(\rho )=C(\rho
^{\ast })$ then we can define a new coherence measure $C^{\prime }(\rho )=%
\frac{1}{2}[C(\rho )+C(\rho ^{\ast })]$ then $C^{\prime }(\rho )=C^{\prime
}(\rho ^{\ast }).$ For Gaussian state $\rho $, since Re$\rho $ in general is
not Gaussian, we consider whether $C_{G}(\rho )-C_{G}(\rho ^{\prime })\geq 0$
where $\rho ^{\prime }$ is a real Gaussian state induced by $\rho $ and $%
C_{G}$ is a coherence measure for Gaussian states.  

Our results revealed some relationships between coherence and imaginarity. The physical
implications and applications of these results are worthy of further investigations.


\section*{Appendix: Some results about symplectic spectrums and Gaussian
states}    \hypertarget{Appendix}{}
\setcounter{equation}{0} \renewcommand\theequation{A\arabic{equation}}

For a Gaussian state $\rho (\overline{X},V),$ the covariance matrix $V$ is a
positive definite real matrix, i.e., $V>0.$ According to Williamson's
theorem \cite{AJM-Williamson-1936}, every positive definite real matrix of even dimension can be
written in diagonal form by a symplectic transformation.
Applying Williamson's theorem to the covariance matrix $V,$ then there
exists a symplectic matrix $M$ such that
\begin{equation}
MVM^{T}=\oplus _{j=1}^{n}\nu _{j}I_{2},   \label{eqA-1}
\end{equation}
where $\nu (V)=\{\nu _{j}(V)\}_{j=1}^{n}$ is (are) called the symplectic
spectrum (eigenvalues) of $V,$ without loss of generality we always assume $%
0<\nu _{j}(V)\leq \nu _{k}(V)$ for $j<k.$ We denote the set of all real
symplectic matrices of $2n$ dimension by
\begin{equation}
\text{Sp}(2n,\mathbb{R})=\{M \big| M\Omega M^{T}=\Omega \}.    \label{eqA-2}
\end{equation}%
Sp$(2n,\mathbb{R})$ forms a group, called symplectic group, and $M\in $Sp$(2n,\mathbb{R})$ if
and only if $M^{T}\in $Sp$(2n,\mathbb{R}).$ The symplectic spectrum $\nu (V)$ is just
the modulus of the $2n$ standard eigenvalues of the matrix $i\Omega V.$ It is easy to check that when
$n=1,$ $V$ has only one symplectic eigenvalue being $\det V.$

The conjugate of the Gaussian state $\rho (\overline{X},V)$ is still a
Gaussian state $\rho ^{\ast }(O\overline{X},OVO).$ Using the facts $%
O^{2}=I_{2n},$ $O\Omega =-\Omega O,$ $O(\oplus _{j=1}^{n}\nu
_{j}I_{2})O=\oplus _{j=1}^{n}\nu _{j}I_{2},$ and Eq. (\ref{eqA-1}), we get
\begin{eqnarray}
(OMO)(OVO)(OMO)^{T} &=&\oplus _{j=1}^{n}\nu _{j}I_{2},   \label{eqA-3}  \\
(OMO)\Omega (OMO)^{T} &=&\Omega .   \label{eqA-4}
\end{eqnarray}
These show that $OMO\in $Sp$(2n,\mathbb{R})$ and the symplectic spectrum of $OVO$ is
still $\{\nu _{j}\}_{j=1}^{n}.$

For the set $\{x_{j}\}_{j=1}^{n}$ of real numbers, define $\{x_{j}^{\uparrow
}\}_{j=1}^{n}$ as the increasing rearrangement of $\{x_{j}\}_{j=1}^{n}$ such
that $x_{1}^{\uparrow }\leq x_{2}^{\uparrow }\leq ...\leq x_{n}^{\uparrow }.$
Two sets $\{x_{j}\}_{j=1}^{n}$ and $\{y_{j}\}_{j=1}^{n}$ of real numbers are
said that $\{x_{j}\}_{j=1}^{n}$ is weakly supermajorized  by $%
\{y_{j}\}_{j=1}^{n}$ (see chapter 1, A.2. in Ref. \cite{book-Marshall-2011}),  denote by
\begin{equation}
\{x_{j}\}_{j=1}^{n}\prec ^{w}\{y_{j}\}_{j=1}^{n},    \label{eqA-5}
\end{equation}
if
\begin{equation}
\sum_{j=1}^{k}x_{j}^{\uparrow }\geq \sum_{j=1}^{k}y_{j}^{\uparrow
},k=1,2,...,n.     \label{eqA-6}
\end{equation}

\begin{Lemma} \label{Lemma-1}
(Theorem 1 in Ref. \cite{PRA-Hiroshima-2006}). Let $A$ and $B$ be $2n\times 2n$
real positive definite matrices $(A=A^{T}>0,B=B^{T}>0).$ Then
\begin{equation}
\nu (A+B)\prec ^{w}\nu (A)+\nu (B).    \label{eqA-7}
\end{equation}
\end{Lemma}

Another interesting result about $\nu (A),$ $\nu (B)$ and $\nu (A+B)
$ is reported in Theorem 2 in Ref. \cite{CMB-Bhatia-2021}.
The following Lemma \label{Lemma-2} is a result of C.1.b. in Chapter 3 of Ref. \cite{book-Marshall-2011}.

\begin{Lemma} \label{Lemma-2}
If $f(x):\mathbb{R}\rightarrow\mathbb{R}$ is concave and increasing, then $\{x_{j}\}_{j=1}^{n}\prec
^{w}\{y_{j}\}_{j=1}^{n}$ implies $\sum_{j=1}^{n}f(x_{j})\geq
\sum_{j=1}^{n}f(y_{j}).$
\end{Lemma}

With Lemma \ref{Lemma-1} and Lemma \ref{Lemma-2}, we have Theorem \ref{Theorem-6} below.

\begin{Theorem} \label{Theorem-6}
For $p\in (0,1)$, Gaussian states $\rho (\overline{X}%
,V) $ and $\sigma (\overline{Y},W),$
it holds that
\begin{equation}
S[p\rho \boxplus (1-p)\sigma ]\geq pS(\rho )+(1-p)S(\sigma ).    \label{eqA-8}
\end{equation}
\end{Theorem}

\emph{Proof.} lemma \ref{Lemma-1} yields
\begin{equation*}
\nu (pV+(1-p)W)\prec ^{w}\nu (pV)+\nu ((1-p)W).
\end{equation*}
From Eqs. (\ref{eq3B-7},\ref{eq3B-9}) and lemma \ref{Lemma-2}, $S(\rho )=\sum_{j=1}^{n}g(\nu _{j}),$ $g(x)$ is
concave and increasing, then
\begin{eqnarray*}
&&S(pV+(1-p)W)  \\
&=&\sum_{j=1}^{n}g[\nu _{j}(pV+(1-p)W)] \\
&\geq &\sum_{j=1}^{n}g[p\nu _{j}(V)+(1-p)\nu _{j}(W)] \\
&\geq &\sum_{j=1}^{n}\{pg[\nu _{j}(V)]+(1-p)g[\nu _{j}(W)]\} \\
&=&pS(\rho )+(1-p)S(\sigma ).
\end{eqnarray*}
$\hfill\square$

Applying Theorem \ref{Theorem-6} to Eq. (\ref{eq3B-3}), we obtain Corollary \ref{Corollary-2} below.

\begin{Corollary} \label{Corollary-2}
For any Gaussian state $\rho ,$ it holds that
\begin{equation}
S(\rho ^{\prime })\geq S(\rho ).  \label{eqA-9}
\end{equation}
\end{Corollary}

The more general result of Theorem \ref{Theorem-6} is the following Corollary \ref{Corollary-3}.

\begin{Corollary}   \label{Corollary-3}
For any probability distribution $\{p_{j}\}_{j}$ and any
Gaussian states $\{\rho _{j}\}_{j},$ it holds that
\begin{equation}
S(\boxplus _{j}p_{j}\rho _{j})\geq \sum_{j}p_{j}S(\rho _{j}),   \label{eqA-10}
\end{equation}
or equivalently,
\begin{equation}
S(\boxplus _{j}p_{j}\rho _{j})-\sum_{j}p_{j}S(\rho _{j})\geq 0,   \label{eqA-11}
\end{equation}
where $\boxplus _{j}p_{j}\rho _{j}=p_{1}\rho _{1}\boxplus p_{2}\rho _{2}....$
\end{Corollary}

Recall that, the concavity of the entropy (see chapter 11 in Ref. \cite{Nielsen-2010-quantum}) says that $S(\sum_{j}p_{j}\rho
_{j})\geq \sum_{j}p_{j}S(\rho _{j})$ for any probability distribution $%
\{p_{j}\}_{j}$ and any quantum states $\{\rho _{j}\}_{j},$ and $%
S(\sum_{j}p_{j}\rho _{j})-\sum_{j}p_{j}S(\rho _{j})$ is called the Holevo
information. Then we can view Eq. (\ref{eqA-10} ) as a counterpart of  $S(\sum_{j}p_{j}\rho
_{j})\geq \sum_{j}p_{j}S(\rho _{j}),$ and view Eq. (\ref{eqA-10} ) as a counterpart of
Holevo information.


%

\end{document}